\documentclass[twocolumn,showpacs,prb]{revtex4}

\usepackage{bm}
\usepackage{epsfig}

\begin{document}

\title{Spin-orbit interaction and the 'metal-insulator' transition \\ observed in two-dimensional hole systems}

\author{L.E.~Golub$^1$}
\email{golub@coherent.ioffe.rssi.ru}
\author{S.~Pedersen$^2$}
\affiliation{$^1$A.F.~Ioffe Physico-Technical Institute, Russian
Academy  of Sciences, 194021 St.~Petersburg, Russia \\
$^2$Chalmers University of Technology, Microelectronic and
Nanoscience, S-412~96, G\"{o}teborg, Sweden}


\begin{abstract}
We present calculations of the spin and phase relaxation rates
in GaAs/AlGaAs $p$-type quantum wells.
These rates are used to derive the temperature dependence of the weak-localization correction
to the conductivity. In $p$-type quantum wells both weak localization and
weak anti-localization are present due to the strong spin-orbit
interaction. When determining the total conductivity correction one also have to include the term due to hole-hole interaction. The magnitude of the latter depends on the ratio between the thermal energy and the Fermi energy, $k_{\rm B}T/E_{\rm F}$ and whether the system can be considered
as ballistic $(k_{\rm B}T \tau_{\rm tr} / \hbar>1)$ or diffusive
($k_{\rm B}T \tau_{\rm tr}/\hbar<1$).
We argue that due to the relatively low Fermi energy and the moderate
mobilities, in the $p$-type systems in question,
the conductivity correction arising from hole-hole interactions is
negligible at the highest temperatures accessible in the experiments.
Hence the 'metal-insulator' transition observed at these relatively high
temperatures could
be caused by interference effects. We compare
our calculations of the weak anti-localization correction with the
experimental results from different independent groups with special emphasis on
the experiments by Simmons et al. We find good agreement between predicted
and observed transistion density $p_{c}$.
\end{abstract}


\maketitle

\section{Introduction}
Quantum phenomena such as weak localization and electron-electron
interaction was for many years considered as a very well exploited field of
condensed matter physics. Among other things it was firmly believed, that
a so-called Metal-Insulator Transition (MIT) would not occur in a two dimensional system | for a
review see Ref.~\onlinecite{lee}.  Hence it came quite unexpectedly when Kravchenko
et al. presented experiments, suggesting the existence of such a MIT in
dilute two-dimensional electron systems.~\cite{krav}
In a series of recent works, similar anomalous temperature
dependence of the conductivity has been observed in a wide range of dilute electron
and hole systems, for an extensive overview see Ref.~\onlinecite{abe}.
The results of these reports are all consistent in the sense, that they support the
original observations of Kravchenko et al.

Several quantum phenomena are known to give rise to metallic
behaviour, so-called weak anti-localization which occurs in the presence of
magnetic impurities, spin-orbit scattering,~\cite{hikami} or when bandstructure-induced spin
relaxation is present.~\cite{knap,JETP,SSC,FTP,simon}
Very recently it has
also been demonstrated theoretically for degenerate systems, that the
electron-electron correction to the conductivity  can undergo a 'metal-insulator' transition,
depending on the value of the Fermi-liquid constant $F^{\sigma}_{0}$.~\cite{zala}
Hence from a fundamental point
of view a metallic phase is indeed possible in 2D.~\cite{boris4,boris}
However in real 2D systems weak anti-localization is usually
disregarded, because of its expected weakness in comparison to the
conductivity corrections due to electron-electron interaction.
In this paper we advocate the idea, that in the dilute $p$-type systems with a moderate mobility, localization due to hole-hole interaction is {\em not}
significant. This is due to the fact, that the
strength of hole-hole interaction induced conductivity correction is
weakened when $k_{\rm B}T$ becomes comparable with the Fermi energy $E_{\rm F}.$~\cite{zala}
Only when $k_{\rm B}T/ E_{\rm F} \ll 1$, a logarithmic or linear hole-hole
correction dominates over the weak localization correction. The systems in question
are, due to their high effective mass, in a
regime where $E_{\rm F} / k_{\rm B}T \approx 1$ at least for the highest range of
experimentally accessible temperatures. This means that the corrections due to hole-hole
interaction can be considered inferior in comparison to e.g. weak anti-localization in the diffusive regime.

We theoretically address a certain group of the dilute GaAs/AlGaAs $p$-type systems which have been experimentally studied in Refs.~\onlinecite{simmons1,simmons2,hanein,yoon} and have in common that the samples have a moderate mobility (in contrast to the hole experiments presented by Mills et al.~\cite{mills} which were performed at very high mobilities). The results of these experiments can be summarized as follows: A transition between an isolating phase at low hole densities and a  metallic phase at high hole densities is observed at a critically hole density $p_{c} \approx 0.3 \dots 1.3\times10^{11}$cm$^{-2}$ | by
isolating  (metallic) phase we mean that the resistivity increases
(decreases) with decreasing temperature.

\begin{table*}[ht] \label{tabel} \caption{The resistance at which
the transition occurs $\rho$, transport relaxation time $\tau_{\rm
tr}$, critical densities $p_{c}$, hole effective mass at the Fermi
level $m^*$, transition temperature between ballistic and
diffusive transport $T^{*}$ and Fermi temperature $T_{\rm F}$
extracted from the studied dilute GaAs/AlGaAs $p$-type quantum
wells.}
\begin{center} \begin{tabular}{lcccccc} \hline
Author & Resistance & Transport & Critical  & Effective & Transition        & Fermi\\
     &          &time       & density   & mass  & temperature   & temperature   \\
\hline
    & $\rho$ & $\tau_{\rm tr}$ & $p_{\rm c}$    & $m^*/m_0$ & $T^{*}$   &$T_{\rm F}$    \\
    &(k$\Omega$) &(ps)   &  $(10^{11}$cm$^{-2} )$    &  &(K)        & (K)\\
 \hline Hanein et al. \cite{hanein}     &5   &8.3   &0.25  &  0.12  & 2.3 & 5.6\\
Simmons et al. \cite{simmons1,simmons2} &10  &2.0   &0.51    & 0.13 & 8.7   & 11.6\\
Yoon et al. \cite{yoon}                  &3  &27.2   &1.27   & 0.16     & 0.5   & 22\\
\hline \end{tabular}
\end{center}
\end{table*}

The rest of this paper we divide in the following sections. In
Section~\ref{4} we discuss the different types of quantum corrections
to the conductivity in weakly disorded 2D systems. In Section~\ref{3}
we derive Cooperon spin relaxation rates performing bandstructure calculations of the highly non-parabolic hole energy spectrum in $p$-type quantum wells.
Finally we present in Section~\ref{6} the
temperature  dependence of the resistivity of a $p$-type quantum well at
different hole densities and compare it with the  performed experiments.

\section{Conductivity corrections in 2D systems}\label{4}
As mentioned in the Introduction, there exists several conductivity
corrections in weakly disordered systems. These quantum corrections can
be divided into two categories | interference effects and hole-hole
interaction effects.

The quantum correction due to inter-particle interactions were recently revised
in the paper by Zala et al.~\cite{zala} Their main result was that, when carefully taking
into account both the exchange and Hartree interaction, the sign of the conductivity
correction could change due to the competition between these two contributions. The sign
of the conductivity correction depends on the value of the interaction constant
$F_{0}^{\sigma}$. As expected they also found that for diffusive systems
$(k_{\rm B}T \tau_{\rm tr} / \hbar<1)$ the temperature correction is logarithmic,~\cite{boris3} whereas for
ballistic systems $(k_{\rm B}T \tau_{\rm tr} / \hbar>1)$ the correction is linear.
However in case where the Fermi energy becomes comparable
with the thermal smearing $k_{\rm B}T / E_{\rm F} \approx 1$, this correction
reduces to a temperature independed constant.

To estimate the value of the Fermi energy we have used
$E_{\rm F}= E(k_{\rm F})$, where $E(k)$ is the hole energy dispersion (its calculation is described in the next Section), and
$k_{\rm F}$ is the Fermi wavevector connected with the density at which the 'metal-insulator' transition occurs,  $p_{\rm c}$, via $k_{\rm F}= \sqrt{2 \pi p_{\rm c}}$. In Table~I we display the calculated
values together with the characteristic temperature $T^{*}=\hbar/k_{\rm B}\tau_{\rm tr}$
which determines the cross-over between ballistic and diffusive transport.
The transport time $\tau_{\rm tr}$ was estimated from the experimental
data by using $\tau_{\rm tr}= m^* \mu / e$, where $\mu$ is the Drude  mobility, which should not to be confused with the somewhat misleading peak mobility $\mu_{peak}$ often given in the experimental reports. The effective hole mass is given by $m^* = \hbar k_{\rm F} / v_{\rm F}$ where the Fermi velocity
\begin{equation}
\label{vF}
v_{\rm F} = \left. {1 \over \hbar} {dE \over dk} \right|_{k=k_{\rm F}} \:.
\end{equation}

The interference effects known as weak localization arise from the
constructive interference of pairs of time inversion symmetric
scattering paths. In non-interacting two-dimensional systems their
contribution to the conductivity can be expressed as~\cite{boris4}
\begin{equation} \label{loc} \Delta\sigma_{wl}(T) =  \frac{e^{2}}{\pi
h}\ln{(\omega_\varphi)}\:.  \end{equation}
Here $\omega_{\varphi} = \tau_{\rm tr}/\tau_{\varphi}$ where
$\tau_{\varphi}$ is the phase coherence time given by \cite{boris4}
\begin{equation} \label{tau_phi}
\frac{1}{\tau_{\varphi}}=\frac{k_{\rm B}T}{\hbar k_{\rm F} l}\ln{(k_{\rm F} l)},
\end{equation}
and $l$ is the mean free
path. To summarize, we see that both interference and hole-hole
interactions can give rise to isolating temperature dependences.

We now turn to the situation where the spin-orbit interaction is
included in
the conductivity corrections. In the case of $p$-type quantum wells, the
conductivity correction has the form~\cite{JETP,SSC}
\begin{equation} \label{dsigma} \Delta \sigma_{wl}(T) =  {e^2 \over 2
\pi h}
\left[  2 \ln{\left(  \omega_\varphi + \omega_\| \right)}  + \ln{\left(
\omega_\varphi + \omega_\perp \right)} - \ln{\left(
\omega_\varphi\right)}
\right] \:, \end{equation}
where $\omega_{\|,\perp}=\tau_{\rm tr}/\tau_{\|,\perp}$ are the
relaxation rates of the Cooperon spin
components parallel and perpendicular to the plane of the
quantum well, respectively.

It is interesting to consider this
expression in the limit of very weak spin-orbit interaction
$\omega_{\perp , \|} \ll 1$ and in the regime where spin-orbit
interaction
is strong $\omega_{\perp , \|} \approx 1$. In the first
case the weak localization correction reduces to Eq.~(\ref{loc}), since
we neglect spin-orbit interaction. In the case of strong spin-orbit
interaction however we find
\begin{equation} \label{aloc}
\Delta\sigma_{wl}(T)=-\frac{e^{2}}{2\pi h}\ln{\left( \omega_{\varphi}
\right)}.  \end{equation}
It is important to note the significant differences
between the expressions (\ref{aloc}) and (\ref{loc}). When introducing
strong spin-orbit interaction, the sign of the conductivity correction
changes,
corresponding to so-called weak anti-localization. Hence if in the regime
of strong
spin-orbit interaction one only considers the interference correction,
the
temperature dependence of the conductivity indeed is metallic.

In experiments it is of cause an interplay of both the hole-hole correction
$\Delta \sigma_{hh}(T)$ and the weak localization correction $\Delta\sigma_{wl}(T)$ which is observed
\begin{equation} \Delta\sigma
(T)= \Delta\sigma_{wl}(T)+ \Delta\sigma_{hh}(T),
\end{equation}
hence the
overall conductivity correction is dominated by the hole-hole interaction when
$k_{\rm B}T \ll E_{\rm F}$.
In situations where $k_{\rm B}T \approx E_{\rm F}$ the effect of hole-hole interaction
however is small in comparison with weak localization or weak anti-localization.
In this regime we find that a metallic temperature dependence of the
conductivity indeed is possible when spin-orbit interactions are present.

\section{Spin and phase relaxation in $p$-type quantum wells} \label{3}
In symmetrical quantum wells, each hole level at a given
wavevector is double-degenerate. The two states, $|1\rangle$ and
$|2\rangle$, connected to each other by the time-inversion
operation, have spin projections on the growth axis equal to $\pm
3/2$ at the bottom of the subband. At finite kinetic energy,
mixing of the heavy- and light-hole states occurs due to strong
spin-orbit interaction present in $p$-type structures. Hence even
under spin-independent scattering, transitions $|1\rangle \to
|2\rangle$ are allowed what leads to relaxation of hole spins. It
was shown~\cite{JETP,SSC} that the Cooperon spin relaxation in
$p$-type quantum wells is anisotropic and the weak localization
correction to conductivity is given by Eq.~(\ref{dsigma}), with
two independent parameters $\omega_\|$ and $\omega_\perp$
determined by the matrix elements of hole scattering~\cite{FTP}
\begin{eqnarray}
\label{omega_par}
\omega_\| = 1 - {\overline{V_{11}^2} \over \overline{|V_{11}|^2} +
\overline{|V_{12}|^2}} \:, \\
\label{omega_perp}
\omega_\perp = {2 \: \overline{|V_{12}|^2} \over \overline{|V_{11}|^2} +
\overline{|V_{12}|^2}} \:.
\end{eqnarray}
The bar here means averaging over the scattering angle of holes at
the Fermi surface. Eqs.~(\ref{omega_par}),~(\ref{omega_perp}) show
the rate $\omega_\perp$ is non-zero mainly due to spin-flip
scattering and $\omega_\|$ is finite because of the change of the
spin-conserving scattering probability with increasing of the
kinetic energy ($\Re e V_{11} \gg \Im m V_{11}, |V_{12}|$).

The value of $\omega_\perp$ is equal to $\tau_{\rm tr} / \tau_s$, where $\tau_s$ is
the hole pseudospin relaxation time, characterizing decay of the hole polarization.
It could be obtained from both optical orientation and weak localization experiments.
The value $\omega_\perp / 2 \tau_{\rm tr} = 1/2\tau_s$ has been calculated in
Ref.~\onlinecite{F&B} for different scattering potentials and quantum well widths.

The parameter $\omega_\|$ is essentially different. It also describes decay of
the zero harmonics of the density matrix, however the latter determines the
Cooperon but not the pseudospin relaxation. Hence $\omega_\|$ can not be obtained from optical
experiments. It can however be extracted from magnetoresistance measurements as it
has been done in Ref.~\onlinecite{simon}.

As seen in Eq.(\ref{tau_phi}) the dephasing rate $\omega_\varphi$
depends only logarithmically on the transport time. Therefore using the
reasonable approximation~\cite{simmons1} $\ln{(k_{\rm F} l)} \approx 2$, we get
an expression for
$\omega_\varphi$ which is independent of scattering parameters:
\begin{equation} \label{omega_phi} \omega_{\varphi}=\frac{2k_{\rm B}T}{\hbar
k_{\rm F} v_{\rm F}} \:. \end{equation}
Here it has been taken into account that the mean free path is connected to the hole velocity at the Fermi level via $l = v_{\rm F} \tau_{\rm tr}$.

\begin{figure}[th]
\epsfxsize=5in \epsfysize=3in \centering{\epsfbox{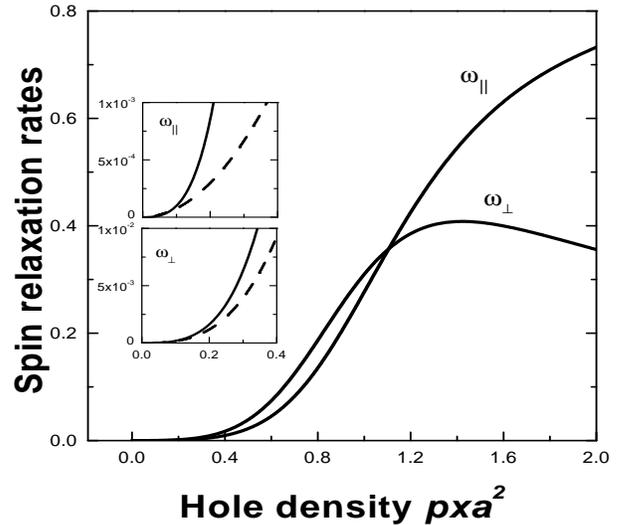}}
\caption{ \label{omegas} The Cooperon spin relaxation rates
$\omega_{\|,\perp}$ in a symmetrical $p$-type quantum well with a
width $a=200$~\AA. Dashed lines in inset show the asymptotic
dependences given by Eq.~(\protect\ref{omega_s_limits}). }
\end{figure}

To obtain the spin relaxation and dephasing rates, we used the
isotropic Luttinger Hamiltonian for calculation of hole states in
a symmetrical quantum well. Assuming the barriers to be infinitely
high, we get the hole energy dispersion in the first
size-quantized subband, $E(k)$. The transcendental equation for
finding $E(k)$ is given, e.g. in Ref.~\onlinecite{MPP}. In this
model, the dispersion is determined only by one parameter, namely
by ratio of the bulk light- and heavy-hole effective masses,
$m_l/m_h$. In calculations we use $m_l/m_h = 0.16$ corresponding
to GaAs-based quantum wells. We take the four-component
wavefunctions in the form proposed in Ref.~\onlinecite{MPP} and
choose the scattering potential as
\begin{equation} \label{V} V({\bm r})= \sum_i V_0 \delta ({\bm r} -
{\bm R}_i) \:, \end{equation} where ${\bm R}_i$ are random
positions of short-range scatters (e.g. impurities). Averaging
squares of matrix elements
in~Eqs.(\ref{omega_par}),~(\ref{omega_perp}) over ${\bm R}_i$, we
get $\omega_\|$ and $\omega_\perp$ as functions of a dimensionless
hole density, $p \times a^2$. Here $a$ is the width of the quantum
well, and the density is connected to the Fermi wavevector as $p =
k_{\rm F}^2 / 2 \pi$.

In Fig.~\ref{omegas} the Cooperon spin relaxation rates
$\omega_{\|,\perp}$ are presented by solid lines. It is important
to note that for short-range scattering~(\ref{V}), $\omega_\|$ and
$\omega_\perp$ are independent of the strength of the potential,
$V_0$. They are determined only by the parameters $m_l/m_h$ and $p
\times a^2$. In this sense the dependences in Fig.~\ref{omegas}
are universal.

In the lowest order in $p \times a^2 \ll 1$, the density
dependences of spin relaxation rates are as
follows~\cite{JETP,SSC}
\begin{equation}
\label{omega_s_limits}
\omega_\| \sim (pa^2)^2 \:,
\hspace{1cm} \omega_\perp \sim (pa^2)^3 \:. \end{equation}
These approximations are plotted by dashed lines in the
insets in Fig.~\ref{omegas}. One
can see that Eq.(\ref{omega_s_limits}) is valid only at extremely low
density. The spin-orbit effects of higher orders start to be important already at $p \geq
0.2/a^2$, and the exact calculation of hole states has to be performed
for obtaining the spin
relaxation rates.

\begin{figure}[th]
\epsfxsize=5in \epsfysize=3in \centering{\epsfbox{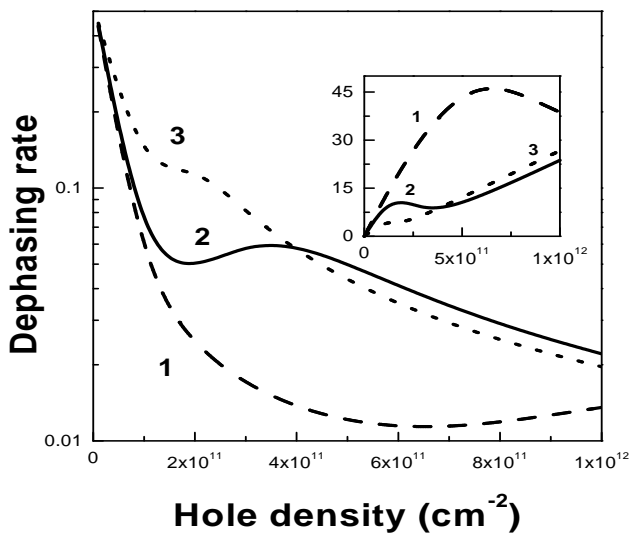}}
\caption{ \label{omegaphi} The dephasing rate $\omega_{\varphi}$
at $T$=1~K at different quantum well widths as a function of hole
density. (1) $a=100$~\AA, (2) $a=200$~\AA, (3) $a=300$~\AA. The
inset displays the density dependence of the parameter $k_{\rm
F}l$ for $\tau_{\rm tr}=2$~ps.}
\end{figure}

The density dependence of the dephasing rate $\omega_\varphi$ is
presented
in Fig.~\ref{omegaphi} for different quantum well widths and $T=1$~K.
Because
of strong non-parabolicity of the hole energy dispersion, the Fermi
velocity~(\ref{vF}) is a non-linear function of $k_{\rm F}$ and, hence, $\omega_\varphi$ changes
with $p$ non-monotonically.

The inset in Fig.~\ref{omegaphi} shows the density dependence of
the parameter $k_{\rm F} l$ at $\tau_{\rm tr} = 2$~ps. It is seen
that $k_{\rm F} l$ also reflects the complicated dependence of the
density of hole states at the Fermi level on concentration.

\section{The Metal-Insulator Transition in $p$-type systems} \label{6}
As seen in Table~I the estimated Fermi temperatures $T_{\rm F}$ is
comparable the highest temperatures obtained in the
experiments in
question. This means that, when considering the quantum correction to
the
conductivity, it is possible to neglect the contribution from hole-hole
interactions. Hence the total resistivity in this temperature regime is
given by
\begin{equation}
\rho(T)={h \over e^2} \left[ {1 \over k_{\rm F}l} - {1 \over (k_{\rm
F}l)^2} {\Delta\sigma_{wl}(T) \over e^2/h} \right],
\end{equation}
where the first term is the classical Drude expression and the second is
the weak localization correction [see Eq.~(\ref{dsigma})].

\begin{figure}
\epsfxsize=5in \epsfysize=3in \centering{\epsfbox{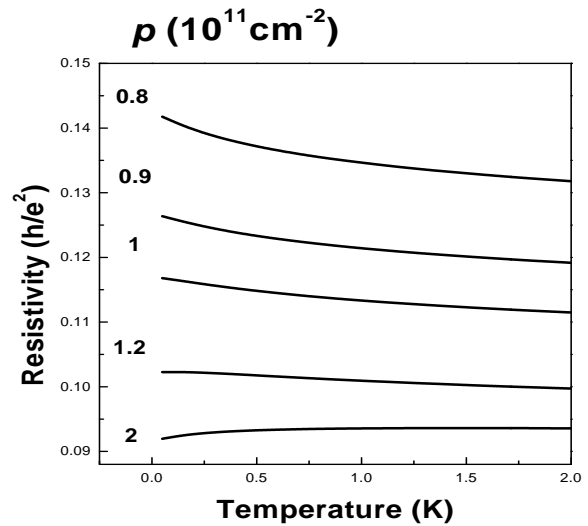}}
\caption{\label{rho} The temperature dependence of the calculated
resistivity for different hole densities in a symmetrical
200~\AA{} wide quantum well. A clear transition between 'metallic'
and 'insulating' behavior occur at approximately
$p=1\times10^{11}$~cm$^{-2}$.}
\end{figure}

In Fig.~\ref{rho} we plot the calculated temperature dependence of the
resistivity
of a 200~\AA{} wide symmetrical quantum well in units of
$h/e^{2}=25.7$~k$\Omega$ for different densities $p$ in the range $0.8
\dots 2 \times 10^{11}$ cm$^{-2}$. To obtain the presented curves we
have used the values of $\omega_{\varphi}$, $\omega_{\perp}$ and
$\omega_{\|}$ calculated in the previous section and a constant
transport time has been taken
$\tau_{\rm tr}=2$~ps. Both the width of the quantum well, $a$, and the
transport time, $\tau_{\rm tr}$, are chosen to be close to the relevant values in the
experiments of Simmons et al.| see Table~I.

From the curves it is seen that at relatively low hole densities the
resistivity increases with decreasing temperatures (isolating behaviour).
At
higher hole densities the opposite situation occurs (metallic behaviour).
Hence we attribute the metal/isolator transition with a weak
anti-localization/localization transition.
This idea has already been analysed in relation to MIT in
electron systems.~\cite{geller,Gornyi} However the spin-orbit
Hamiltonian in $p$-type systems is significantly different from the linear
in the wavevector term in electron systems, therefore MIT in hole systems can not be described by the
theory Ref.~\onlinecite{geller}.

The theoretically observed transition occurs due to the density dependence of the spin scattering. The critical density where the
cross-over between the two different temperature dependences occurs is
approximately $p_c \approx 1\times 10^{11}$cm$^{-2}$,
corresponding to $p \times a^{2} \approx 0.4$. When comparing with the critical hole densities found in the
experimental
reports, we find good agreement between the observed and the predicted
transition density | see Table~I. Especially it is worthwhile to note
the
good agreement between theory and the experiments by Simmons et
al.~\cite{simmons1,simmons2} which were performed on a 200~\AA{} wide
symmetrical quantum well in accordance with our assumptions.

\begin{figure}[t]
\epsfxsize=5in \epsfysize=3.2in \centering{\epsfbox{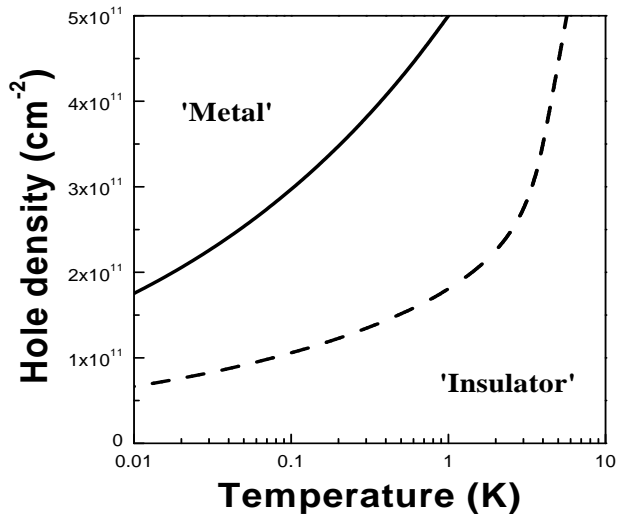}}
\caption{ \label{Phase_diagramm} The phase diagram of a hole gas
confined in a 200~\AA{} (solid line) and 100~\AA{} (dashed line)
wide quantum well. }
\end{figure}

The calculated temperature dependences furthermore display a weak
non-monotonic features close to the critical
density | see the curves for $p=1.2\times 10^{11}$cm$^{-2}$ and
$p=2\times 10^{11}$cm$^{-2}$. Hence close to the
critical density the temperature dependence of the resistivity displays
a
local maxima. This behaviour is in agreement with experiments and was
especially pronounced in the reports by Hanein et al.~\cite{hanein},
Yoon et al.~\cite{yoon}
and Mills et al.~\cite{mills}. This peculiarity of the interference
correction is included into Eq.~(\ref{dsigma}). Indeed, for very low
temperatures the conductivity correction is metallic, but for high
temperatures the system is insulating. For a given density this local
maximum takes place when $\partial \Delta\sigma_{wl} / \partial T = 0$,
i.e. at
\begin{equation}
\label{zero_slope}
{\omega_\varphi} = \sqrt{ \left( {\omega_\perp \over 4} \right)^2 +
{\omega_\| \omega_\perp \over 2 }}  - {\omega_\perp \over 4 } \:.
\end{equation}
This equation connects the density with temperature at the maximum.
The corresponding dependence $p_{max}(T)$ shown in
Fig.~\ref{Phase_diagramm} represents the 'phase diagram' of the diluted
hole gas.

\section{Conclusion} \label{7}
We have calculated the density dependence of the spin and phase relaxation rates in
a $p$-type quantum well. By applying these results to the
theory of weak-localization, we find that it is possible to explain the
so-called 'metal-insulator' transition which is observed in a range
of 2D hole systems at temperatures $k_{\rm B}T \approx E_{\rm F}$ and at moderate mobilities.
This transition is explained as a crossover between
weak localization and weak anti-localization due to the strong density
dependence of the spin and phase relaxation rates.
We have made a direct comparison between our results and the
experimental work by Simmons et al.~\cite{simmons1,simmons2} and find a good agreement for the observed critical
densities $p_{c}$. For the sake of completeness we have in Table~I listed the values found for several other relevant experiments.

We underline that our theory only is applicable in the regime
$k_{\rm B}T \approx E_{\rm F}$ and hence do not address situations where the system
is degenerate, which normally is the case for electrons.

Finally we would like to encourage experimentalists to apply optical
techniques and hence shine light on the connection between the metal-insulator
transition and spin-relaxation. We also attract their attention to the non-monotonous
temperature dependences of resistivity and the corresponding 'phase diagram' of the 2D
hole system.

\subsection*{Acknowledgements}
This work is financially supported by the RFBR, by the Programmes
of Russian Ministry of Science and Presidium of RAS, and by INTAS.


\end{document}